\documentclass[prl,twocolumn]{revtex4}

\usepackage{amsthm,amssymb,amsmath}
\usepackage{graphicx}
\usepackage{epsfig}

\begin{document}
\title{Large contributions to dark matter annihilation from  three-body final states}
\author{Carlos E. Yaguna}
\email{carlos.yaguna@uam.es}
\affiliation{Departamento de F\'{i}sica Te\'orica C-XI, Universidad Aut\'onoma de Madrid, Cantoblanco, E-28049 Madrid, Spain}
\affiliation{Instituto de F\'{i}sica Te\'orica UAM-CSIC, Universidad Aut\'onoma de Madrid, Cantoblanco, E-28049 Madrid, Spain}

\begin{abstract}
The annihilation rate of dark matter particles plays a crucial role in dark matter studies, for it determines their relic density and their indirect detection signal. In this paper, we show that this annihilation rate can receive large additional contributions from three-body final states consisting of a real and a virtual massive particle, such as $WW^*$ ($\to Wf\bar f'$) and $t\bar t^*$ ($\to tW\bar b$). We consider  two specific examples, from the singlet model and the MSSM, and  find that, due to the new three-body final state contributions, the prediction for the relic density may decrease by more than a factor two, whereas  the present dark matter annihilation rate gets enhanced by up to two orders of magnitude. Some of the implications of these results are briefly discussed.

\end{abstract}
\maketitle

\section{Introduction}
Dark matter annihilations have  two different and equally important roles in the phenomenology of dark matter models. On the one hand, the annihilation of   dark matter particles in the early Universe dictates, via the Boltzmann equation,  the expected relic density and, consequently, the viable parameter space of dark matter models.  This parameter space is usually obtained by requiring, on top of accelerator and precision bounds, that the relic density  be consistent with the observed dark matter density \cite{Dunkley:2008ie}. Hence, an accurate computation of the dark matter annihilation cross section is almost a prerequisite for dark matter studies.  In addition, the present annihilation rate of dark matter particles determines the indirect detection signatures of dark matter. Thus, to find the expected fluxes in gamma rays, neutrinos, and antimatter generated by dark matter annihilations,  we must first compute properly the cross section for the annihilation of dark matter particles.

Up to now, most studies have only consider dark matter annihilations  into either  two-body final states, or  three-body final states with a massless gauge boson \cite{Bergstrom:1989jr}, such as $q\bar q g$ or $f\bar f\gamma$, and have implicitly assumed that they dominate the total annihilation cross section. There exists, however, another class of three-body final states that may have an important impact on dark matter annihilations. They are final states consisting of a real and a virtual massive particle, such as $WW^*$ ($Wf\bar f'$) and $t\bar t^*$  ($t W\bar b$), and are the main subject of this paper. These three-body final states have not been considered in much detail in the literature. In fact, neither DarkSUSY \cite{Gondolo:2004sc} nor micrOMEGAs \cite{Belanger:2006is}, two sophisticated programs commonly used to study the phenomenology of dark matter models, include any of these three-body final states in their calculations. To our knowledge, the only works where such final states have been considered are  \cite{Chen:1998dp} and \cite{Hosotani:2009jk}. In \cite{Chen:1998dp}, Chen and Kamionkowski computed the cross section for neutralino annihilation into the three-body final states $WW^{*}$ and $t\bar t^*$ and showed that, in certain regions of the parameter space, they can enhance the neutrino signal expected from neutralino annihilations. In that work, which has remained largely unnoticed over all these years, they did not study the possible effect of three-body final states on the neutralino relic abundance; they simply stated that such effect should be negligible. In \cite{Hosotani:2009jk}, Hosotani, Ko, and Tanaka considered  stable higgs  bosons as cold dark matter candidates in the context of gauge-higgs unification models. They found, using a semi-analytic approximation,  that the inclusion of the three-body final state $WW^{*}$ modifies significantly the prediction of the relic density for dark matter masses below the $W$ threshold. In this paper, we will update and improve these previous results in several ways. We will  accurately compute the effect of three-body final states on the annihilation rate of dark matter particles and on their relic abundance in two well-motivated extensions of the standard model: the MSSM and the singlet scalar model. In both cases, we find that the inclusion of three-body final states considerably affects the predicted relic density and the indirect detection signal of dark matter particles. We thus demonstrate that  three-body final states consisting of a real and a virtual massive particle cannot be generally neglected in dark matter annihilations.

\section{Three-body final states}
Ordinarily, particle physics processes such as decays and annihilations are dominated --provided that they are open-- by two-body final states. Three-body final states are typically suppressed with respect to them by one additional coupling and one more propagator. Exceptions to this behaviour are not uncommon though. The decay of the higgs boson in the standard model constitute a particularly striking example of the importance of three-body final states. An intermediate mass higgs boson, $135$ GeV $\lesssim m_h\lesssim 160$ GeV, decays mainly into  a three body final state consisting of a real $W$ boson and a virtual one $(W^*)$ that in turn decays into a fermion-antifermion pair, $h\to WW^*\to Wf\bar f'$ \cite{Djouadi:2005gi}. This three-body final state accounts for more than $10\%$ of the branching for higgs masses above $116$ GeV and dominates over $b\bar b$ in a wide region below the $W^+W^-$ threshold. Thus, to properly compute the higgs decay width and its branching ratios, two-body decays are not enough; three-body final states must necessarily be taken into account. By the same token, we will see that, in some cases, to accurately compute the dark matter relic density and its indirect detection signals it is mandatory to include three-body final states.

Besides illustrating the need to go beyond two-body final states, the analogy with the decays of the higgs boson also allows us  to foresee the  conditions under which three-body final states are expected to be relevant in dark matter annihilations. Notice that, in the higgs case, the decay $h\to W^+W^-$ becomes dominant once it is open, and it is below that threshold that the three-body decay (into $WW^*$) is important. For dark matter annihilations, accordingly, three-body final states could be relevant below the threshold of a standard model particle, $X$, if the annihilation into $X\bar X$ becomes dominant for $m_{DM}>m_X$, where $m_{DM}$ denotes the mass of the dark matter particle. Given that, in most dark matter models, $m_{DM}\sim 100~$GeV (WIMPs), $X$ could be  a $W$ (or $Z$) boson, a higgs boson, or a top-quark. Hence, three-body final states might be relevant for dark matter masses below $M_W$, $m_h$, or $m_t$. 
To satisfy the other condition observed in the higgs case, we need that the annihilation into $W^+W^-$, $hh$ or $t\bar t$ become dominant above the respective threshold. Thus, whether three-body final states turn out to be relevant or not will depend on the specific particle physics model considered and on the region of the parameter space examined.

We will use two examples,  singlet scalar dark matter and  neutralino dark matter in the MSSM, to illustrate the relevance of three-body final states in dark matter annihilations. In both cases we find regions in the parameter space where the inclusion of three-body final states modifies in a significant way the dark matter phenomenology of these models.  

\section{Singlet scalar dark matter}

\begin{figure}[b]
\begin{center} 
\includegraphics[scale=0.5]{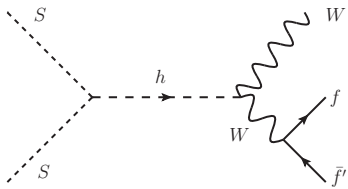}
\caption{One of the diagrams with three-body final states that enhances the dark matter annihilation rate in the singlet model.\label{fig:diasinglet}}
\end{center}
\end{figure}
The singlet scalar model \cite{McDonald:1993ex,Yaguna:2008hd} is one of the simplest extensions of the standard model that may explain the dark mattter. It includes one additional scalar field, $S$, that is neutral under the standard model gauge group and odd under a new discrete symmetry $Z_2$ ($S\to -S$). This symmetry enforces the stability of $S$ and renders it a suitable dark matter candidate. This simple model introduces only two additional parameters to the standard model: the singlet mass, $m_S$, and the coupling between the singlets and the higgs, $\lambda$. In addition, the higgs mass ($m_h$), a standard model parameter, also enters into dark matter calculations. It has been shown elsewhere, see e.g. \cite{Yaguna:2008hd}, that the singlet scalar model can indeed explain the observed dark matter density and that it can be tested through direct and indirect detection in future experiments. 

\begin{figure}[tb]
\begin{center} 
\includegraphics[scale=0.3]{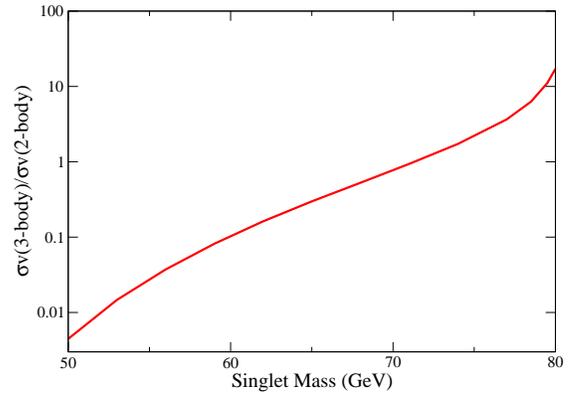}
\caption{The ratio between the three-body and the two-body annihilation rate in the singlet scalar model of dark matter.\label{fig:singlet}}
\end{center}
\end{figure}

In this model, singlets annihilate mainly through s-channel higgs boson exchange into standard model fermions and gauge bosons. Among the two-body final states, the main annihilation channels are $b\bar b$ for $m_S<M_W$ and   $W^+W^-$ for $m_S>M_W$. It is a situation analogous to that of higgs boson decays mentioned previously. Henceforth, it is likely that within the mass range
\begin{equation}
50~\textrm{GeV}\lesssim m_S\lesssim M_W,
\end{equation}
the annihilation cross section is actually dominated by the three-body final state $WW^*$ ($W^+W^{-*}+W^{+*}W^-$). Figure \ref{fig:diasinglet} shows the Feynmann diagram corresponding to the annihilation process into this three-body final state, $SS\to WW^*\to Wf\bar f'$, in the singlet model. Notice that there are $18$ different final states contributing to that diagram. How important is this new contribution?

Figure \ref{fig:singlet} shows, as a function of the singlet mass,  the ratio between the annihilation rate, $\sigma v$,  into the three-body final state $WW^*$ and that into two-body final states, which is dominated by the $b\bar b$ contribution. The figure was obtained for $\lambda=10^{-2}$, $m_h=120$ GeV, and small $v$ ($\approx 10^{-3}$), typical for dark matter particles annihilating in the Galactic halo. To compute the cross sections we used the CalcHEP \cite{calchep} and micrOMEGAs \cite{Belanger:2006is} packages. Even though $\sigma v(\mathrm{3\text{-}body})$ and $\sigma v(\mathrm{2\text{-}body})$ are both resonant for $m_S\sim 60$ GeV, their ratio has a smooth non-resonant behavior, as observed in the figure. Moreover,  because $\lambda$ and $m_h$ affect both cross sections in the same way, the line shown in figure \ref{fig:singlet} is actually independent of these parameters. In other words, it is a generic prediction of the singlet model.

Notice from the figure that already for $m_S\sim 60~$GeV the three-body final state is not negligible, accounting for about $10\%$ of the cross section. For $m_S\sim 70~$GeV, the two-body and the three-body cross sections are of the same order. Finally, between $m_S\sim 70~$GeV and $m_S\sim M_W$, the singlet annihilation cross section is clearly dominated by the three-body final state. In fact, close to $M_W$ the cross section into $WW^*$ is more than 10 times larger than the total cross section into two-body final states. Hence, for singlet masses between $60$ GeV and $M_W$, the three-body final state cannot be neglected in indirect detection studies, for it modifies considerably the total annihilation rate and the branching fractions into standard model particles, giving rise to different yields in photons, neutrinos, and antimatter. 

\begin{figure}[tb]
\begin{center} 
\includegraphics[scale=0.3]{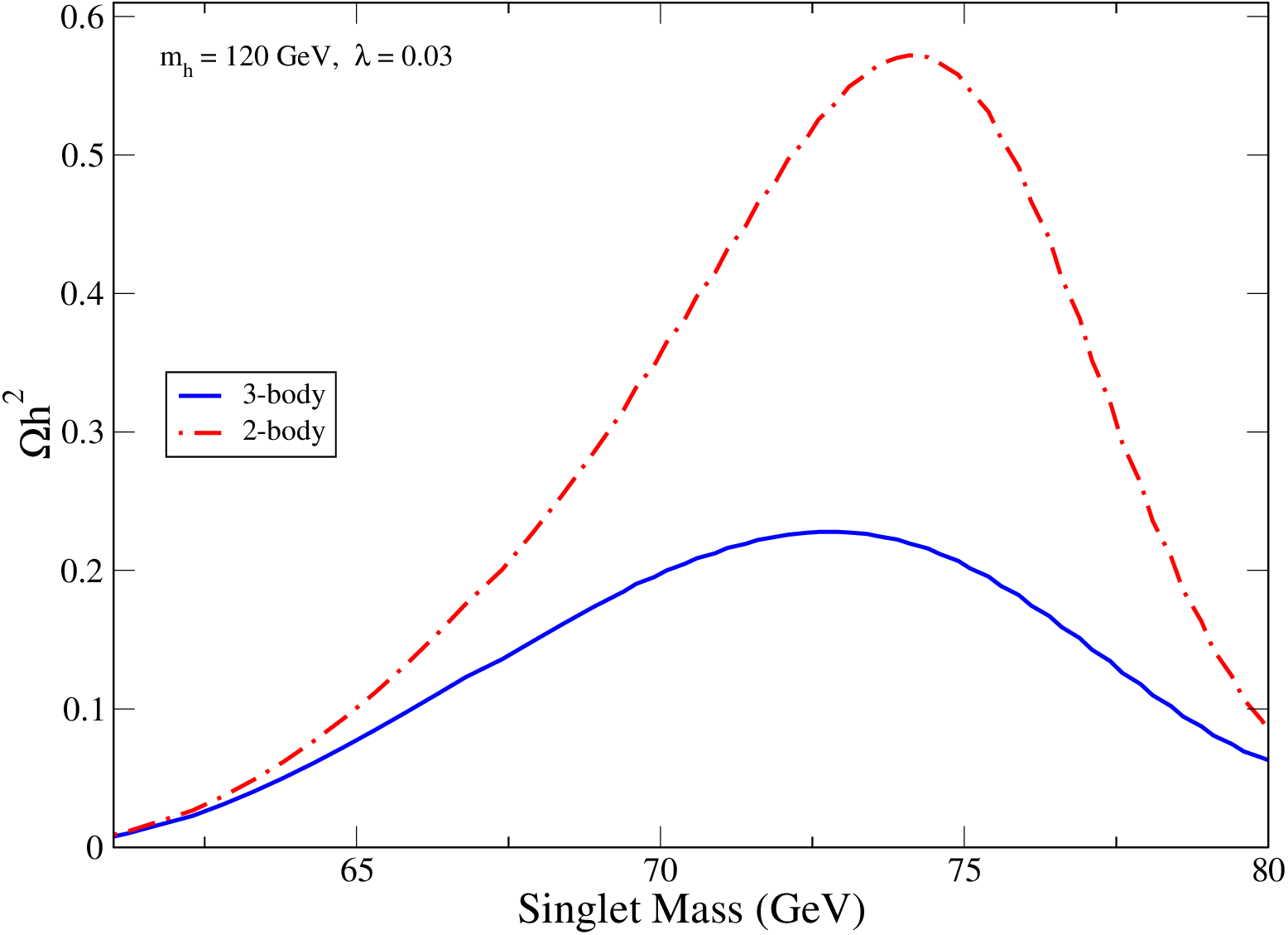}
\caption{The ratio between the relic density obtained including the three-body final state and the one predicted for two-body final states only. The parameter $\lambda$ was set to $0.03$. \label{fig:rd120}}
\end{center}
\end{figure}

\begin{figure}[tb]
\begin{center} 
\includegraphics[scale=0.3]{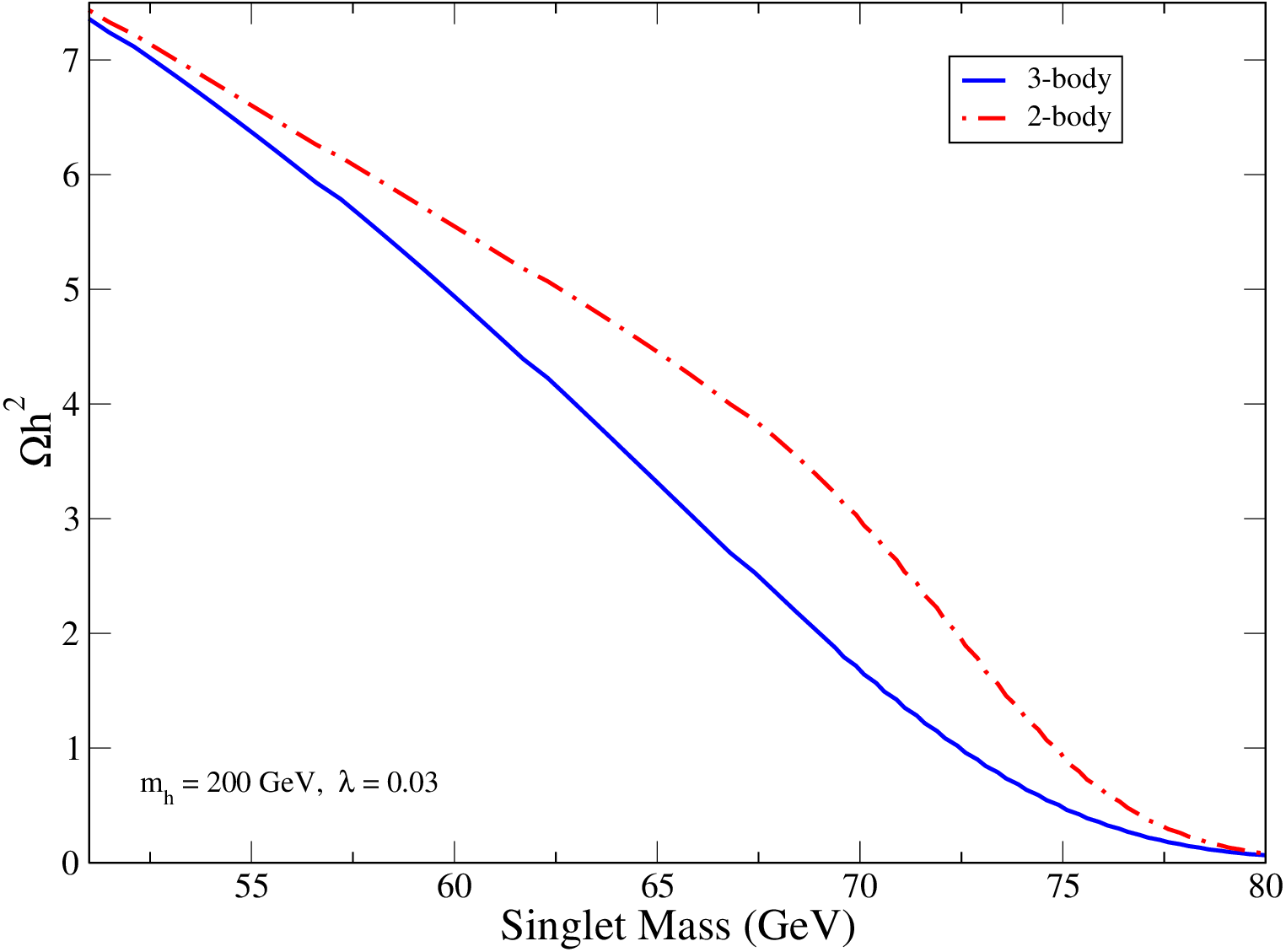}
\caption{The ratio between the relic density obtained including the three-body final state and the one predicted for two-body final states only. The parameter $\lambda$ was set to $0.03$. \label{fig:rd200}}
\end{center}
\end{figure}

\begin{figure}[tb]
\begin{center} 
\includegraphics[scale=0.3]{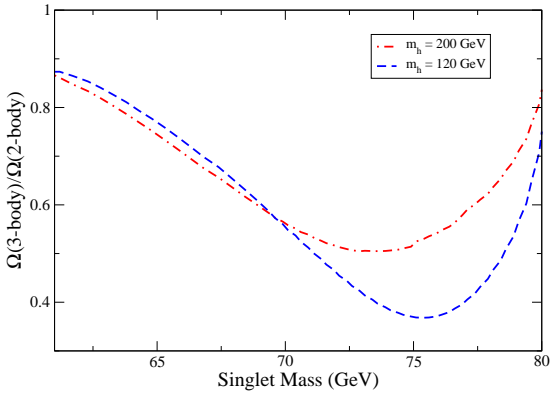}
\caption{The ratio between the relic density obtained including the three-body final state and the one predicted for two-body final states only. The parameter $\lambda$ was set to $0.03$. \label{fig:rdsinglet}}
\end{center}
\end{figure}

Due to the higher dark matter velocity in the early Universe and to the thermal averaging of the cross section, the effect of the three-body final state on the singlet relic density is expected to be smaller. It is significant nonetheless, as illustrated by figures \ref{fig:rd120}-\ref{fig:rdsinglet}. Figure \ref{fig:rd120}  compares the correct relic density (solid line), obtained including  two- and three-body  final states, with the predicted relic density for two-body final states only (dash-dotted line), as usually computed in the literature. For that figure we set $m_h=120$ GeV and $\lambda=0.03$. To obtain these results we again used CalcHEP, to compute the three-body contribution,  and micrOMEGAs, to calculate the two-body part and to solve the Boltzmann equation.
Both lines have the same generic behavior. The relic density  initially increases as the singlet mass departs from the higgs resonance (located at $m_S\sim 60$ GeV), it reaches a maximum, and it then decreases in the vicinity of the $W$-threshold. The value of the correct relic density, however, may differ significantly from that predicted for annihilation into two-body final states. At $m_S\sim 75$ GeV, for instance, the two-body approximation overestimates the cross section by more than a factor $2$. 

A similar effect is present also for larger higgs masses, as illustrated in figure \ref{fig:rd200}. In this case the relic density is a decreasing function of the singlet mass. Again, we observe that the two-body approximation overestimates the relic density for singlet masses between $50$ GeV and $M_W$. A better way to visualize the effect of the three-body final state on the relic density is by displaying the ratio of the correct relic density (two and three-body) to the two-body relic density, as done in figure \ref{fig:rdsinglet}. Notice that the ratio is  significantly smaller than $1$ over a wide mass range and that it can decrease even below $0.4$, a large deviation from the two-body result. From this figure we learn that in the singlet model the correct relic density could be significantly smaller than the relic density obtained for two-body final states.

A detailed analysis of the resulting new parameter space of the singlet model, and of the implications of three-body final states for its direct and indirect detection prospects will be done in a future work. Next, we will see the relevance of three-body final states for neutralino dark matter.

\section{Neutralino dark matter}

\begin{figure}[b]
\begin{center} 
\includegraphics[scale=0.6]{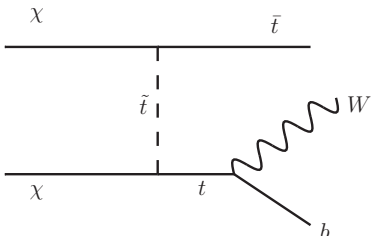}
\caption{One of the diagrams with three-body final states that enhances the neutralino annihilation rate in the MSSM.\label{fig:diamssm}}
\end{center}
\end{figure}

For neutralino dark matter the most promising region where three-body final states may play an important role is below the top-quark threshold. That is, for neutralino masses in the range
\begin{equation}
130~\textrm{GeV}\lesssim m_\chi\lesssim m_t\,.
\label{eq:range}
\end{equation}
In this mass interval, neutralinos can annihilate into a $t\bar t^*$ pair followed by the decay of the virtual $\bar t$ into $W^-\bar b$ ($\chi\chi\to t\bar t^*\to tW\bar b$).  One of the diagrams that contribute to this process in the MSSM is illustrated in figure \ref{fig:diamssm}. For bino-like neutralinos, the dominant two-body annihilation channel in such mass interval is typically $b\bar b$. Can the three-body final state dominate over $b\bar b$?
\begin{table}[b]
\begin{tabular}{|c|c|c|c|c|c|c}
\hline
Parameter & $M_2$, $M_3$, $m_{\tilde \ell}$,$\mu$ &$m_{\tilde q}$ &$A_t$& $M_A$ &$\tan\beta$\\
\hline
Value & $1$ TeV &$500$ GeV & $1.4$ TeV &$1.5$ TeV &5\\
\hline

\end{tabular}
\caption{Supersymmetric parameters used. $M_1$ was allowed to vary freely.\label{table:mssm}}
\end{table}

To ensure that diagrams with three-body final states, such as figure \ref{fig:diamssm}, are not suppressed, we will focus on a region of the  parameter space featuring  a relatively light stop. Table \ref{table:mssm} shows the  supersymmetric parameters, defined at low energy, that we consider in the following. They give rise to models compatible with present bounds from accelerator and precision data. The lightest stop, which is lighter than all other squarks thanks to the  non-zero trilinear coupling $A_t$, has a mass of about $260$ GeV. The neutralino mass is determined by $M_1$ and is left as a free parameter to be varied in the range (\ref{eq:range}). The resulting lightest neutralino is always bino-like. 

\begin{figure}[tb]
\begin{center} 
\includegraphics[scale=0.3]{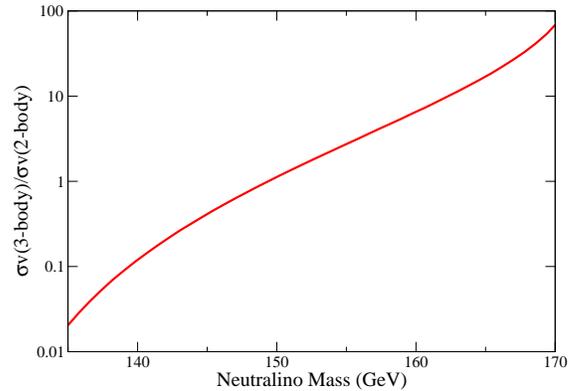}
\caption{The ratio between the three-body and the two-body annihilation rate in the supersymmetric model defined by  Table \ref{table:mssm}.\label{fig:mssm}}
\end{center}
\end{figure}

Figure \ref{fig:mssm} shows, as a function of the neutralino mass,   the ratio between the cross section into the three-body final state $t\bar t^*$ and that one into  two-body final states. Since the latter is almost constant over the neutralino mass range shown, the variation observed in the figure is entirely due to the  cross section into the three-body final state. As before, we took $v\approx 10^{-3}$ and used micrOMEGAs and CalcHEP to compute the cross sections. Notice from the figure that at $m_\chi\sim 140~$GeV the three-body final state already reaches $10\%$ of the total annihilation cross section. For $m_\chi\sim 150~$GeV, the three-body final state contribute to the cross section as much as the two-body states. For larger masses, up to the top threshold, the neutralino annihilation cross section is entirely dominated by the three-body final state. In that region, the two-body final state approximation commonly used in previous works  would fail badly, underestimating the neutralino annihilation cross section and misjudging the resulting final states. As a consequence, neither the normalization nor the spectrum of the $\gamma$, $\nu$, $e^+$ and $\bar p$ generated in neutralino annihilations would be correct.

\begin{figure}[tb]
\begin{center} 
\includegraphics[scale=0.3]{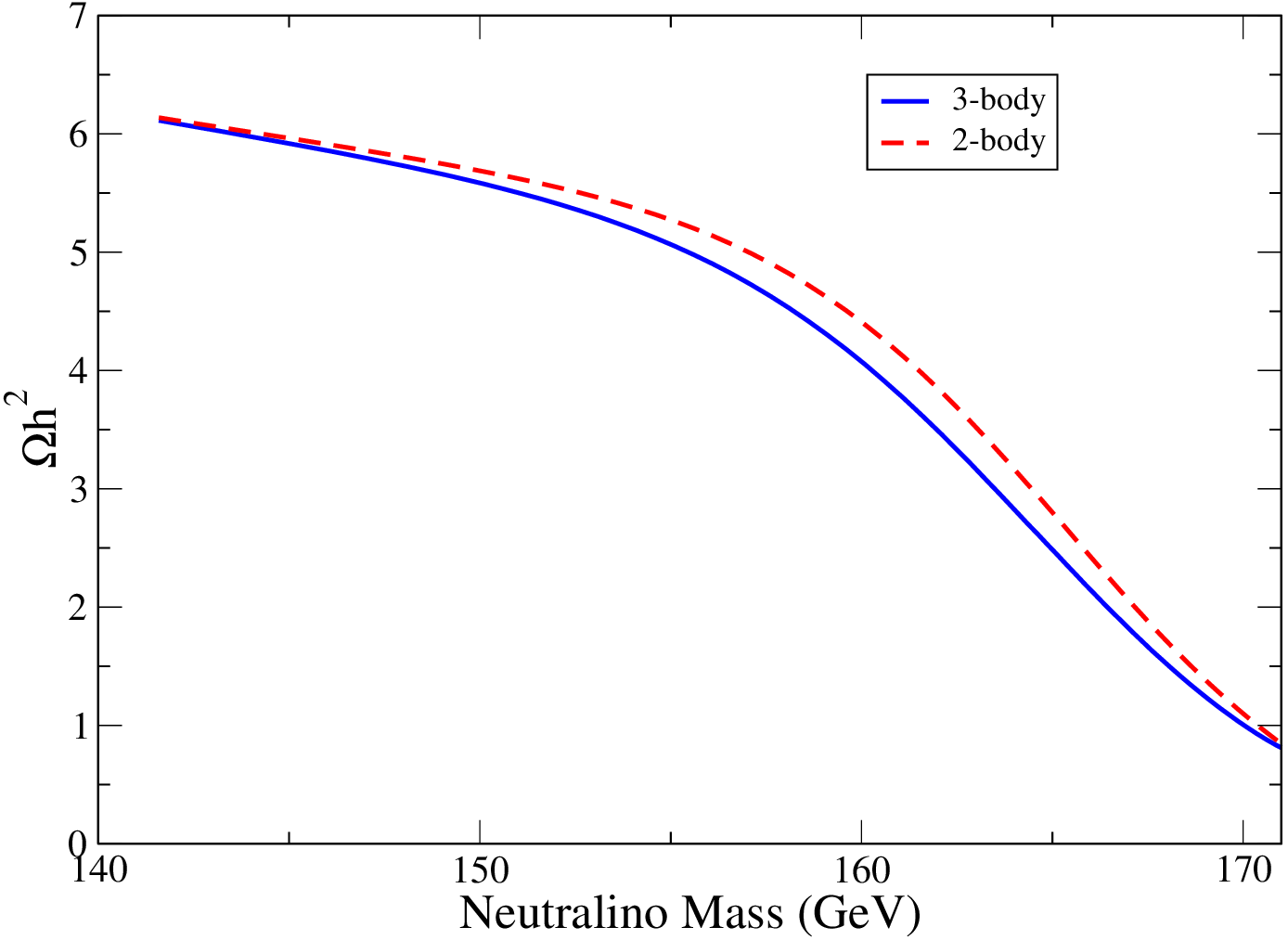}
\caption{The ratio between the relic density obtained including the three-body final state and the one predicted for two-body final states only. \label{fig:relicmssm}}
\end{center}
\end{figure}

\begin{figure}[tb]
\begin{center} 
\includegraphics[scale=0.3]{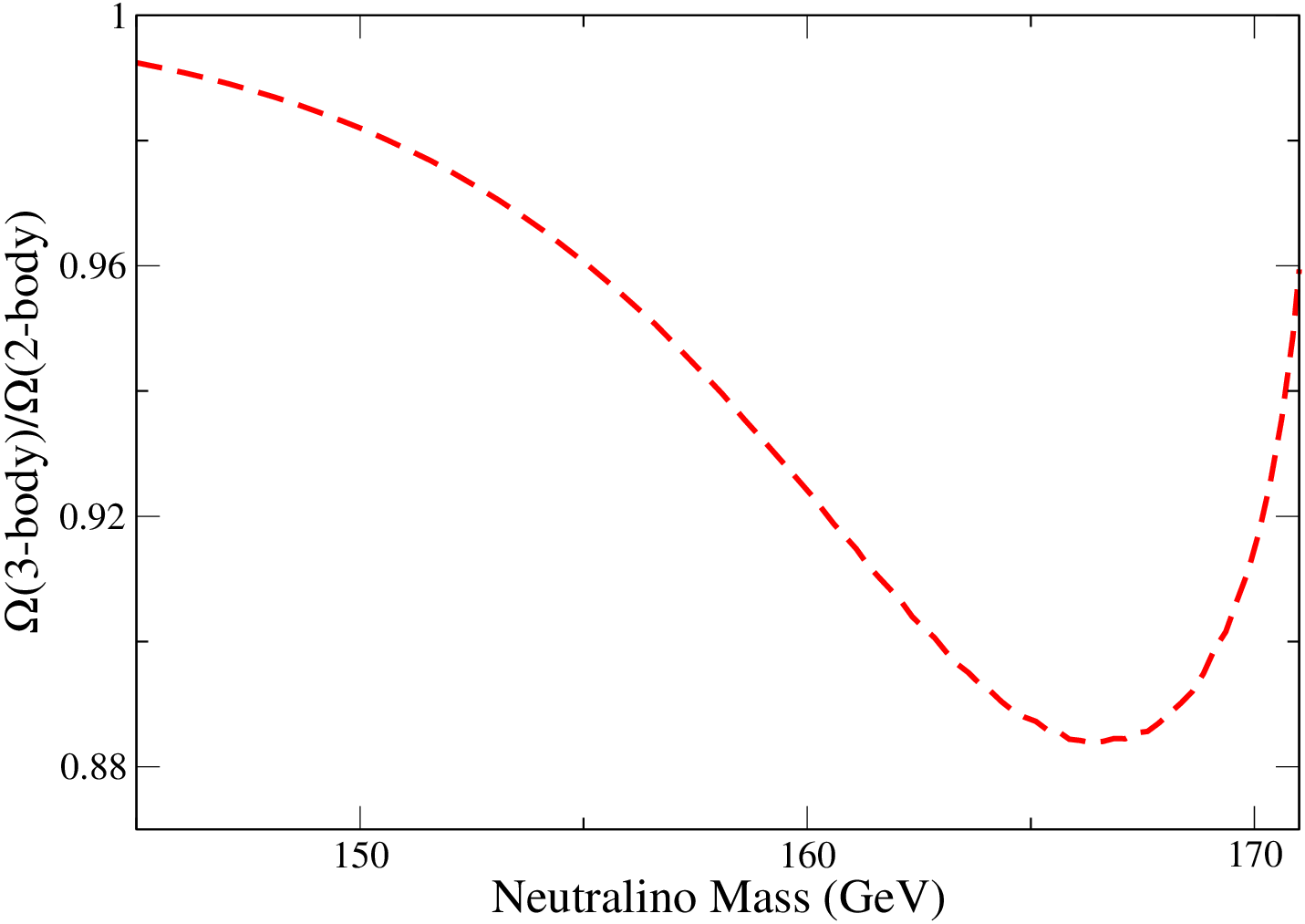}
\caption{The ratio between the relic density obtained including the three-body final state and the one predicted for two-body final states only. \label{fig:rdmssm7}}
\end{center}
\end{figure}

The effect of the three-body final state on the neutralino relic density is illustrated in figures \ref{fig:relicmssm} and \ref{fig:rdmssm7}. In the calculation, coannihilation effects between the neutralino and the lightest stop are automatically taken into account. They are not important though, because the mass difference between them is significant. Notice that the correct neutralino abundance, calculated including the three-body final state, could be more than $10\%$ smaller than that obtained taking into account only two-body final states, as usually done in the literature.

In view of these results,  the claim that DarkSUSY computes the neutralino relic density, for any MSSM model,  with an accuracy of $1\%$ \cite{Gondolo:2005we} no longer holds. To reach that accuracy, the three-body final states  studied in this work must necessarily be included in the calculation.

In any case, notice that, in contrast with  the singlet model studied in the previous section, the large effect due to three-body final states below the top threshold is not a generic feature of the MSSM. It is only within certain configurations, for specific sets of parameters, that the effect becomes important. It must be also  said  that there certainly exist additional corrections not included in our analysis, such as one-loop corrections to the cross sections and toponium resonance effects below the top threshold, that could also modify the neutralino annihilation cross section  and the predicted neutralino relic density in the MSSM.  

The most significant feature brought about by dark matter annihilation into the three-body final state $t\bar t^*$ in supersymmetric models is that, as we have seen, it may give rise to large corrections to the neutralino  annihilation cross section and to the relic density of neutralino dark matter.

\section{Conclusions}
We have shown that  three-body final states consisting of a real and a virtual massive particle may  play an important role in dark matter annihilations. Two well-motivated scenarios were used to illustrate this effect: the singlet scalar and the MSSM.  In both cases, we demonstrated that these new contributions can significantly modify the prediction for the relic density of dark matter and for its indirect detection signatures. In a future work we will study in more detail the implications of these results.

\begin{acknowledgments}
It is a pleasure to  thank Laura for useful discussions and for bringing reference \cite{Hosotani:2009jk} to my attention. I am supported by the \emph{Juan de la Cierva} program of the Ministerio de Educacion y Ciencia of Spain. I acknowledge additional support from  the Comunidad de Madrid under grant HEPHACOS S2009/ESP-1473 and from the MICINN Consolider-Ingenio 2010 Programme under grant MULTIDARK CSD2009-00064.
\end{acknowledgments}

\end{document}